\documentclass[aip,graphicx]{revtex4-1}
\usepackage[cp1251]{inputenc}
\usepackage[T2A]{fontenc}
\usepackage[english]{babel}
\usepackage{amssymb,latexsym,amsmath,amscd}
\usepackage{graphicx,color,framed}
\usepackage{xcolor}
\usepackage{siunitx} 
\begin{document}
\selectlanguage{english}
\title{On gyration radius distributions of star-like macromolecules}
\author{\firstname{Yury A.} \surname{Budkov}}
\email[]{ybudkov@hse.ru}
\affiliation{School of Applied Mathematics, HSE University, Tallinskaya St. 34, 123458 Moscow, Russia}
\affiliation{G.A. Krestov Institute of Solution Chemistry of the Russian Academy of Sciences, 153045, Akademicheskaya St. 1, Ivanovo, Russia}
\author{Andrei L. Kolesnikov}
\email[]{kolesnikov@inc.uni-leipzig.de}
\affiliation{Institut f\"ur Nichtklassische Chemie e.V., Permoserstr. 15, 04318 Leipzig, Germany}

\begin{abstract}
Using the path integral approach, we obtain the characteristic functions of the gyration radius distributions for Gaussian star and Gaussian rosette macromolecules. We derive the analytical expressions for cumulants of both distributions. Applying the steepest descent method, we estimate the probability distribution functions of the gyration radius in the limit of a large number of star and rosette arms in two limiting regimes: for strongly expanded and strongly collapsed macromolecules. We show that in both cases, in the regime of a large gyration radius relative to its mean-square value, the probability distribution functions can be described by the Gaussian functions. In the shrunk macromolecule regime, both distribution functions tend to zero faster than any power of the gyration radius. Based on the asymptotic behavior of the distribution functions and the behavior of statistical dispersions, we demonstrate that the probability distribution function for the rosette is more densely localized near its maximum than that for the star polymer. We construct the interpolation formula for the gyration radius distribution of the Gaussian star macromolecule which can help to take into account the conformational entropy of the flexible star macromolecules within the Flory-type mean-field theories.

\end{abstract}
\maketitle
\section{Introduction}
Polymers with a branched structure are very promising compounds for modern materials science and engineering. Branched macromolecules are usually used as elementary structural units of a vast range of functional polymer materials, such as liquid gels and aerogels, elastomers, melts, thermoplastics, {\sl etc.} High functionality of branched polymers opens new opportunities for their use in drug delivery, robotics, antibacterial and antifouling surfaces, food industry and others applications. In biological processes, such as stabilization of globular proteins or transcriptional regulation of genes, an important role is played by multiple loop formation in branched macromolecules. That is why it is of fundamental interest to study the conformational properties of complex polymer architectures \cite{haydukivska2020universal,douglas1990characterization}.

In order to understand the role of ring constituents in compactification of branched polymers, the authors of paper \cite{blavatska2015conformational} considered a model of a star-like polymer, where one part of the arms were linear Gaussian polymers, while the rest were ring ones. The authors studied the conformational behavior of star-like Gaussian macromolecules containing a total of $f_1$ ring arms (petals) and $f_2$ linear chains (branches). Within the framework of the continuous model of a Gaussian macromolecule, they applied the path integral approach to estimate the gyration radius and asphericity of the typical conformation as functions of the $f_1$ and $f_2$ parameters. In particular, their results reveal the extent of anisotropy of star-like topologies relative to the rosette structures of the same total molecular weight. Moreover, their analytical results quantitatively confirm the compactification (decrease in the effective size) of multiple loop polymer structures as compared with structures containing only linear segments. The authors of paper \cite{haydukivska2020universal} determined the mean-square gyration radius and the average hydrodynamic radius of the star-like Gaussian polymer, and estimated their ratio with its dependence on the numbers of petals and branches. Moreover, they obtained quantitative estimate of the compactification degree of these polymers with an increasing number of petals as compared to linear or star-shaped molecules of the same total molecular weight \cite{haydukivska2020universal}.

Despite the fact that the behavior of the mean-square gyration radius of branched macromolecules has been extensively studied \cite{blavatska2015conformational,haydukivska2020universal}, the behavior of the probability distribution function (PDF) of the gyration radius of star-like macromolecules remains unclear. In particular, it is important to understand how the replacement of the linear branches with ring ones can qualitatively change the PDF behavior. In this paper, we attempt to answer this question, considering two limiting examples of the branched polymer: a Gaussian star polymer with $f$ identical branches (see Fig. 1a) and a Gaussian rosette (Fig. 1b) with $f$ petals. Each arm is a Gaussian linear polymer for the star macromolecule and a Gaussian ring polymer -- for the rosette. In the framework of the path integral approach proposed in our previous paper \cite{Budkov2016}, we for the first time obtain analytical expressions for characteristic functions of the gyration radius distributions and then estimate the asymptotics of the corresponding PDF for star and rosette macromolecules in the limit of a large number of arms. 

\begin{figure}[h]
\centering
\includegraphics[width=1.0\linewidth]{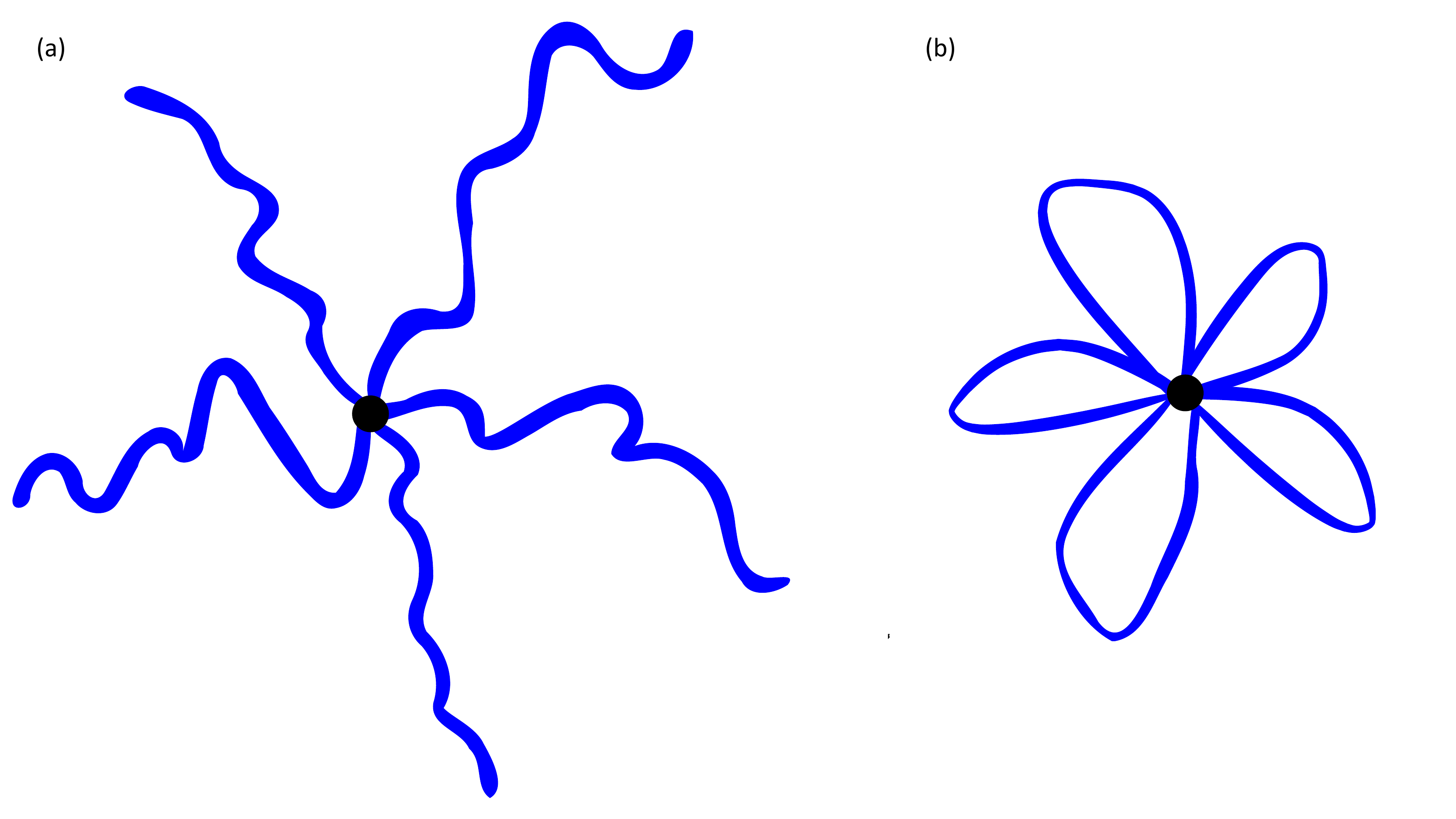}
\caption{Schematic representation of a Gaussian star (a) and a Gaussian rosette (b).}
\label{fig:corr_pore} 
\end{figure}

\section{Statement of the problem}
The probability distribution function (PDF) of the gyration radius of star-like Gaussian macromolecules with $f$ linear or ring arms (see Fig. 1(a,b)) can be written as the following path integral
\begin{equation}
P(R_{g}^2)=\int\frac{\mathcal{D}\bold{r}_{1}}{Z_{0}^{(1)}}..\int\frac{\mathcal{D}\bold{r}_{f}}{Z_{0}^{(f)}}e^{-\frac{3}{2b^2}\sum\limits_{i=1}^{f}\int\limits_{0}^{N}d\tau\dot{\bold{r}}_{i}^2(\tau)}\delta\left(R_{g}^2-\frac{1}{2N^2f^2}\sum\limits_{i,j=1}^{f}\int\limits_{0}^{N}\!\!\int\limits_{0}^{N}d\tau_1 d\tau_2\left(\bold{r}_{i}(\tau_1)-\bold{r}_{j}(\tau_2)\right)^2\right),
\end{equation}
where 
\begin{equation}
Z_{0}^{(i)}=\int{\mathcal{D}\bold{r}_{i}} \exp\left[-\frac{3}{2b^2}\int\limits_{0}^{N}d\tau\dot{\bold{r}}_{i}^2(\tau)\right]
\end{equation}
are the normalization constants of the Gaussian measures; $N$ is the polymerization degree and $b$ is the Kuhn length. Using the Fourier-representation of the delta-function
\begin{equation}
\delta(x)=\int\limits_{-\infty}^{\infty}\frac{d\xi}{2\pi}e^{-i\xi x},
\end{equation}
we obtain
\begin{equation}
P(R_{g}^2)=\int\limits_{-\infty}^{\infty}\frac{d\xi}{2\pi}e^{-i\xi R_{g}^2}K(\xi),
\end{equation}
where
\begin{equation}
K(\xi)=\int\frac{\mathcal{D}\bold{r}_{1}}{Z_{0}^{(1)}}..\int\frac{\mathcal{D}\bold{r}_{f}}{Z_{0}^{(f)}}e^{-\frac{3}{2Nb^2}\sum\limits_{i=1}^{f}\int\limits_{0}^{1}ds\dot{\bold{r}}_{i}^2(s)+\frac{i\xi}{2f^2}\sum\limits_{i,j=1}^{f}\int\limits_{0}^{1}\!\!\int\limits_{0}^{1}ds_1 ds_2\left(\bold{r}_{i}(s_1)-\bold{r}_{j}(s_2)\right)^2}
\end{equation}
is the characteristic function \cite{gnedenko2018theory}, where we turned to the variables $s=\tau/N \in [0,1]$. Our aim is to calculate the characteristic function $K(\xi)$ as the Gaussian path integral and then to evaluate the PDF by the steepest descent method \cite{lavrentiev1987methods}. As is well known, the characteristic function can be related to the moments $\mu_{k}$ and cumulants $\chi_k$ by the expansions
\begin{equation}
K(\xi)=1+\sum\limits_{k=1}^{\infty}\frac{\mu_{k}}{k!}(i\xi)^{k},
\end{equation}
\begin{equation}
\label{Ln_K}
\ln K(\xi)=\sum\limits_{k=1}^{\infty}\frac{\chi_{k}}{k!}(i\xi)^{k},
\end{equation}
so that the moments $\mu_k$ and cumulants $\chi_{k}$ are related to each other by the expressions
\begin{equation}
\label{moments}
\mu_{1}=\chi_{1},~\mu_{2}=\chi_{2}+\mu_{1}^2,..
\end{equation}

\section{Characteristic functions}
In this section, we demonstrate a derivation of the analytical expressions for the characteristic functions of star and rosette Gaussian polymers within the path integral approach.
\subsection{Star polymer}
Here, we consider the case of a star-like Gaussian polymer --  a set of $f$ identical Gaussian chains grafted to one point (see, Fig. 1a). Using the equality
\begin{equation}
e^{-\frac{\bold{b}^2}{4a}}=\left(\frac{a}{\pi}\right)^{3/2}\int d\bold{x}e^{-a\bold{x}^2+i\bold{b}\bold{x}},
\end{equation} 
the characteristic function can be rewritten as follows
\begin{equation}
\nonumber
K(\xi)=\int..\int\prod\limits_{i=1}^{f}\frac{\mathcal{D}\bold{r}_{i}}{Z_{0}^{(i)}}e^{-\frac{3}{2Nb^2}\int\limits_{0}^{1}ds\left(\dot{\bold{r}}_{i}^2(s)-\omega^2\bold{r}_{i}^2(s)\right)}\exp\left[-\frac{i\xi}{f^2}\left(\sum\limits_{i=1}^{f}\int\limits_{0}^{1}ds\bold{r}_{i}(s)\right)^2\right]=
\end{equation}
\begin{equation}
\label{K}
\left(\frac{f^2}{4\pi i\xi}\right)^{3/2}\int d\bold{x} e^{\frac{if^2\bold{x}^2}{4\xi}}\left[\int\frac{\mathcal{D}\bold{r}}{Z_{0}}e^{-S[\bold{r}]}\right]^{f},
\end{equation}
where the auxiliary functional
\begin{equation}
\label{S}
S[\bold{r}]=\frac{3}{2Nb^2}\int\limits_{0}^{1}ds\left(\dot{\bold{r}}^2(s)-\omega^2\bold{r}^2(s)\right)-i\bold{x}\int_{0}^{1}ds\bold{r}(s),
\end{equation}
and notation $\omega^2=2Nb^2i\xi/3f$ are introduced. Then it is necessary to calculate the path integral in eq. (\ref{K}). Using the saddle-point method \cite{Budkov2016,paradezhenko2020gaussian,zinn2010path}, we obtain
\begin{equation}
\label{path}
\int\frac{\mathcal{D}\bold{r}}{Z_{0}}e^{-S[\bold{r}]}=\left(\frac{3}{2\pi Nb^2}\right)^{3/2}\left(\frac{\omega}{\sin{\omega}}\right)^{3/2}\int d\bold{R}e^{-S[\bold{v}_{SP}]},
\end{equation}
where
\begin{equation}
\bold{v}_{SP}(s)=-\frac{i\bold{x}Nb^2}{3\omega^2}\left(1-\cos{\omega s}-\frac{\sin{\omega s}(1-\cos{\omega})}{\sin{\omega}}\right)+\frac{\bold{R}\sin{\omega s}}{\sin{\omega}}
\end{equation}
is the solution of the Euler-Lagrange equation $\delta S[\bold{r}]/\delta \bold{r}(s)=0$ with the boundary conditions $\bold{r}(0)=0$ and $\bold{r}(1)=\bold{R}$. Taking into account that 
\begin{equation}
S[\bold{v}_{SP}]=\frac{3R^2\omega}{2Nb^2\tan{\omega}}-\frac{i\bold{R}\bold{x}\tan{\frac{\omega}{2}}}{\omega}+\frac{i\bold{x}^2f}{4\xi}-\frac{i\bold{x}^2f}{2\xi}\frac{\tan{\frac{\omega}{2}}}{\omega},
\end{equation}
after integration over the end point of the chain, $\bold{R}$, in eq. (\ref{path}), and subsequent integration over the auxiliary variable, $\bold{x}$, we arrive at the new analytical expression for the characteristic function
\begin{equation}
\label{char_func}
K(\xi)=\left(\frac{\omega}{\sin{\omega}}\right)^{3/2}\left(\frac{1}{\cos{\omega}}\right)^{3(f-1)/2}.
\end{equation}
Note that for a linear Gaussian chain ($f=1$), eq. (\ref{char_func}) transforms into the well-known characteristic function $K(\xi)=(\omega/\sin{\omega})^{3/2}$ that was first obtained by Fixman \cite{fixman1962radius} within the Wang-Uhlenbeck method \cite{yamakawa1971modern}. The logarithm of the characteristic function can be expanded into the following power series
\begin{equation}
\ln K(\xi)=\sum\limits_{k=1}^{\infty}\left(\frac{2}{3}\right)^{k-1}\frac{2^{2k-1}}{k(2k)!}(1+(f-1)(2^{2k}-1))B_{2k-1}\left(\frac{Nb^2}{f}\right)^{k}(i\xi)^{k},
\end{equation}
so that the cumulants (see, eq. (\ref{Ln_K})) are determined by
\begin{equation}
\label{chi}
\chi_{k}=\left(\frac{2}{3}\right)^{k-1}\frac{2^{2k-1}(k-1)!}{(2k)!}(1+(f-1)(2^{2k}-1))B_{2k-1}\left(\frac{Nb^2}{f}\right)^{k}, 
\end{equation}
where we use the following power series
\begin{equation}
\ln\frac{\omega}{\sin{\omega}}=\sum\limits_{k=1}^{\infty}\frac{2^{2k-1}B_{2k-1}}{k(2k)!}\omega^{2k},
\end{equation}
\begin{equation}
\ln\cos{\omega}=\sum\limits_{k=1}^{\infty}\frac{2^{2k-1}B_{2k-1}(1-2^{2k})}{k(2k)!}\omega^{2k},
\end{equation}
with the Bernoulli numbers $B_{1}=1/6$, $B_{3}=1/30$, $B_{5}=1/42$,...
In particular, we obtain the well-known expression \cite{zimm1949dimensions,blavatska2015conformational}
\begin{equation}
\left<R_{g}^2\right>=\mu_{1}=\chi_{1}=\frac{3f-2}{f}\frac{Nb^2}{6}, \label{eq:star_first_moment}
\end{equation}
and the second moment
\begin{equation}
\label{mu2}
\left<R_{g}^4\right>=\mu_{2}=\chi_{2}+\mu_{1}^2=\frac{(Nb^2)^2}{540}\frac{135f^2-120f+4}{f^2}.
\end{equation}
In a similar way, using eq. (\ref{chi}), we can calculate the higher moments of the distribution.

\subsection{Rosette}
Now let us calculate the characteristic function for a Gaussian rosette with $f$ petals (Fig. 1b). Performing the same calculations as in the previous subsection, we obtain
\begin{equation}
\label{char_3}
K(\xi)=\left(\frac{f^2}{4\pi i\xi}\right)^{3/2}\int d\bold{x} e^{\frac{if^2\bold{x}^2}{4\xi}}\left[\int\frac{\mathcal{D}\bold{r}}{Z_{0}}e^{-S[\bold{r}]}\right]^{f},  
\end{equation}
where the random vector-function $\bold{r}(s)$ satisfies the following boundary conditions $\bold{r}(0)=\bold{r}(1)=0$ (the rosette center is placed at the origin) and the $S[\bold{r}]$ functional is determined by eq.(\ref{S}). Expanding the random vector-function, $\bold{r}(s)$, into a Fourier series
\begin{equation}
\bold{r}(s)=\sqrt{2}\sum\limits_{n=1}^{\infty}{\bf\rho}_n \sin{\pi ns},
\end{equation}
we turn to the integration over the Fourier components\cite{Budkov2016}, $\bold{\rho}_{n}$, i.e.
\begin{equation}
\nonumber
\int\frac{\mathcal{D}\bold{r}}{Z_{0}}e^{-S[\bold{r}]}=
\frac{\int..\int\prod\limits_{n=1}^{\infty}d\bold{\rho}_{n}\exp\left[-\frac{3}{2Nb^2}\sum\limits_{n=1}^{\infty}(\pi^2n^2-\omega^2)\bold{\rho}_n^2+\frac{2i\sqrt{2}\bold{x}}{\pi}\sum\limits_{k=1}^{\infty}\frac{\bold{\rho}_{2k-1}}{2k-1}\right]}{\int..\int\prod\limits_{n=1}^{\infty}d\bold{\rho}_{n}\exp\left[-\frac{3}{2Nb^2}\sum\limits_{n=1}^{\infty}\pi^2n^2\bold{\rho}_n^2\right]}=   
\end{equation}
\begin{equation}
\nonumber
\left(\prod\limits_{n=1}^{\infty}\frac{\pi^2n^2}{\pi^2n^2-\omega^2}\right)^{3/2}\exp\left[-\frac{4Nb^2\bold{x}^2}{3\pi^2}\sum\limits_{k=1}^{\infty}\frac{1}{(2k-1)^2(\pi^2(2k-1)^2-\omega^2)}\right]=
\end{equation}
\begin{equation}
\label{fourier}
\left(\frac{\omega}{\sin\omega}\right)^{3/2}\exp\left[\frac{i\bold{x}^2f}{4\xi}\left(\frac{2}{\omega}\tan{\frac{\omega}{2}-1}\right)\right].
\end{equation}
Therefore, substituting eq. (\ref{fourier}) for eq. (\ref{char_3}), after the integration over variable $\bold{x}$, we arrive at the new analytical expressions for the characteristic function for the Gaussian rosette
\begin{equation}
\label{char_2}
K(\xi)=\left(\frac{\omega}{\sin{\omega}}\right)^{3(f-1)/2}\left(\frac{\omega}{2\sin{\frac{\omega}{2}}}\right)^{3}.
\end{equation}
Eq. (\ref{char_2}) allows us to obtain analytical expressions for cumulants. Indeed, the logarithm of the characteristic function takes the form
\begin{equation}
\ln{K(\xi)}=\sum\limits_{k=1}^{\infty}\frac{B_{2k-1}}{k(2k)!}\left(\frac{2}{3}\right)^{k-1}\left(1+2^{2k-1}(f-1)\right)\left(\frac{Nb^2}{f}\right)^{k}(i\xi)^{k},
\end{equation}
so that the cumulants can be calculated by the following analytical expression:
\begin{equation}
\label{chi2}
\chi_{k}=\frac{(k-1)!}{(2k)!}\left(\frac{2}{3}\right)^{k-1}\left(1+2^{2k-1}(f-1)\right)B_{2k-1}\left(\frac{Nb^2}{f}\right)^{k}.
\end{equation}
Using eqs. (\ref{moments}), we obtain the moments:
\begin{equation}
\label{moments}
\mu_{1}=\left<R_{g}^2\right>=\frac{Nb^2}{12}\frac{2f-1}{f},~\mu_{2}=\left<R_{g}^4\right>=\frac{60f^2-44f+1}{f^2}\frac{N^2b^4}{2160},...
\end{equation}
Note that the above expression for the first moment, $\mu_{1}$, was obtained recently in paper \cite{blavatska2015conformational}.

\section{Probability distribution functions}
In this section, we apply the steepest descent method to study the asymptotic behavior of the PDF of the Gaussian star-like macromolecules discussed above \cite{lavrentiev1987methods}.
\subsection{Star polymer}
For the star polymer, the PDF can be rewritten in the form
\begin{equation}
P(R_g^2)=\int\limits_{-\infty}^{+\infty}\frac{d\xi}{2\pi}\exp\left[-W(\xi)\right],
\end{equation}
where we introduce the following auxiliary function
\begin{equation}
W(\xi)=\frac{(3f-2)\alpha^2\omega^2(\xi)}{4}-\frac{3}{2}\ln{\frac{\omega(\xi)}{\sin{\omega(\xi)}}}+\frac{3(f-1)}{2}\ln{\cos{\omega(\xi)}}
\end{equation}
and expansion factor $\alpha=R_{g}/\left<R_{g}^2\right>^{1/2}$. In order to estimate the PDF, we use the steepest descent method \cite{lavrentiev1987methods}. At first, we have to find the saddle-point, $\xi_{0}$, from the solution of the saddle-point equation
\begin{equation}
W^{\prime}(\xi_0)=\frac{3}{2\xi_0}\left(\frac{(3f-2)\alpha^2\omega_0^2}{6}-\frac{1}{2}\left(1-\frac{\omega_0}{\tan{\omega_0}}\right)-\frac{1}{2}(f-1)\omega_0\tan{\omega_0}\right)=0,
\end{equation}
where $\omega_0=\omega(\xi_0)$. Note that the contour of integration must be deformed along the steepest descent line. In the vicinity of $\xi_0$ we have
\begin{equation}
W(\xi)=W(\xi_0)+\frac{1}{2}W^{\prime\prime}(\xi_0)(\xi-\xi_0)^2+..,
\end{equation}
where the second derivative takes the form
\begin{equation}
W^{\prime\prime}(\xi_0)=\frac{3}{8\xi_0^2}\left(2-\frac{\omega_0}{\tan{\omega_0}}-\frac{\omega_0^2}{\sin^2{\omega_0}}+(f-1)\left(\omega_0\tan{\omega_0}-\frac{\omega_0}{\cos^2{\omega_0}}\right)\right).
\end{equation}

In what follows, we will consider only the case of a large number of arms ($f\gg 1$) for two limiting regimes -- $\alpha\gg 1$ and $\alpha\ll 1$. In this $"$thermodynamic$"$ limit, the steepest descent method must yield the exact asymptotic behavior of the PDF. The saddle-point equation at $f\gg 1$ takes the following simplified form
\begin{equation}
\frac{\tan{\omega_0}}{\omega_0}=\alpha^2,
\end{equation}
whereas the second derivative is
\begin{equation}
W^{\prime\prime}(\xi_0)\simeq \frac{3f}{8\xi_0^2}\left(\omega_0\tan{\omega_0}-\frac{\omega_0^2}{\cos^2{\omega_0}}\right).  
\end{equation}
For the highly expanded polymer star ($\alpha\gg 1$), the saddle-point is $\omega_0\simeq \frac{\pi}{2}-\frac{2}{\pi \alpha^2}$, which yields
\begin{equation}
W(\xi_0)\simeq \frac{3f\pi^2\alpha^2}{16}-\frac{3f}{2}\ln{\frac{\pi\alpha^2}{2}}-\frac{3f}{2},~W^{\prime\prime}(\xi_0)\simeq \frac{\alpha^4N^2b^4}{6f}. 
\end{equation}
Thus, calculating the integral, we obtain
\begin{equation}
P(R_{g}^2)\simeq \left(\frac{3f}{\pi}\right)^{1/2}\left(\frac{e\pi}{2}\right)^{3f/2}\frac{\alpha^{3f}}{Nb^2}\exp\left[-\frac{3\pi^2f\alpha^2}{16}\right].    
\end{equation}
In the opposite case, $\alpha\ll 1$, the saddle-point is $\omega_0\simeq{i}/{\alpha^2}$. In this case, we have
\begin{equation}
W(\xi_0)\simeq \frac{3f}{4\alpha^2}-\frac{3f}{2}\ln{2},~W^{\prime\prime}(\xi_0)\simeq \frac{\alpha^6 N^2b^4}{6f},
\end{equation}
and the PDF acquires the following form
\begin{equation}
P(R_g^2)=\left(\frac{3f}{\pi}\right)^{1/2}\frac{2^{3f/2}}{\alpha^3Nb^2}\exp\left[-\frac{3f}{4\alpha^2}\right].
\end{equation}
Collecting together the above results, we eventually obtain the following limiting laws for the PDF
\begin{equation}
\label{PDF1}
P(R_g^2)\simeq
\begin{cases}
\left(\frac{3f}{\pi}\right)^{1/2}\left(\frac{e\pi}{2}\right)^{3f/2}\frac{\alpha^{3f}}{Nb^2}\exp\left[-\frac{3\pi^2f\alpha^2}{16}\right], &\alpha\gg 1\,\\
\left(\frac{3f}{\pi}\right)^{1/2}\frac{2^{3f/2}}{\alpha^3Nb^2}\exp\left[-\frac{3f}{4\alpha^2}\right], &\alpha\ll 1.
\end{cases}
\end{equation}
It is worth noting that the PDF can be described by the Gaussian function only at $\alpha\gg 1$ (PDF has a Gaussian tail). At small $\alpha$, the PDF goes to zero faster than any positive power of $\alpha$. The obtained asymptotic relations (\ref{PDF1}) allow us to estimate the limiting behavior of the conformational entropy, $S(R_g^2)=\ln{P(R_{g}^2)}$, of the Gaussian star polymer in the limit $f\gg 1$. Thus, we have
\begin{equation}
\label{entropy}
{S(R_g^2)}/{f}\simeq
\begin{cases}
-\frac{3\pi^2\alpha^2}{16}+3\ln{\alpha}+\frac{3}{2}\ln{\frac{\pi e}{2}}, &\alpha\gg 1\,\\
-\frac{3}{4\alpha^2}+\frac{3}{2}\ln{2}, &\alpha\ll 1,
\end{cases}
\end{equation}
where we kept only the terms that are proportional to $f$. We would like to note that the limiting relations (\ref{entropy}) can be used for a theoretical description of the conformational behavior of star macromolecules in solvent media in the same manner as the conformational entropy of the linear Gaussian chain \cite{Budkov2016,fixman1962radius} can be used within the Flory-type mean-field theories \cite{Budkov2016,budkov2018models}. 

\subsection{Rosette}
In a similar way, for the Gaussian rosette, the PDF can be written in the form
\begin{equation}
P(R_g^2)=\int\limits_{-\infty}^{+\infty}\frac{d\xi}{2\pi}\exp\left[-W(\xi)\right],
\end{equation}
where we introduce the following auxiliary function
\begin{equation}
W(\xi)=\frac{(2f-1)\alpha^2\omega^2(\xi)}{8}-3\ln{\frac{\omega(\xi)}{2\sin{\frac{\omega(\xi)}{2}}}}-\frac{3(f-1)}{2}\ln{\frac{\omega(\xi)}{\sin{\omega(\xi)}}},
\end{equation}
with the expansion factor, $\alpha^2=R_{g}^2/\left<R_{g}^2\right>$. The saddle-point equation is
\begin{equation}
W^{\prime}(\xi_0)=\frac{3}{2\xi_{0}}\left(\frac{(2f-1)\alpha^2\omega_0^2}{12}+\frac{\omega_0}{2\tan{\frac{\omega_0}{2}}}-1-\frac{f-1}{2}\left(1-\frac{\omega_0}{\tan{\omega_0}}\right)\right),  
\end{equation}
where $\omega_0=\omega(\xi_0)$. The second derivative is
\begin{equation}
W^{\prime\prime}(\xi_0)=\frac{3}{4\xi_0^2}\left(f+1-\frac{\omega_0}{2\tan{\frac{\omega_0}{2}}}-\frac{\omega_0^2}{4\sin^2{\frac{\omega_0}{2}}}-\frac{1}{2}(f-1)\frac{\omega_0}{\sin{\omega_0}}\left(\cos{\omega_0}+\frac{\omega_0}{\sin{\omega_0}}\right)\right).
\end{equation}
As for the star polymer, we consider only the case of a large number of arms ($f\gg 1$). The saddle-point equation in this case is simplified to
\begin{equation}
\frac{3}{\omega_0^2}\left(1-\frac{\omega_0}{\tan{\omega_0}}\right)={\alpha^2}.
\end{equation}
In the regime of a highly expanded macromolecule ($\alpha\gg 1$), the solution of the saddle-point equation is $\omega_0\simeq\pi-3/\pi\alpha^2$, which yields
\begin{equation}
W(\xi_0)\simeq\frac{\pi^2\alpha^2f}{4}-\frac{3f}{2}\ln{\frac{\pi^2e\alpha^2}{3}},~W^{\prime\prime}(\xi_0)\simeq \frac{\alpha^4N^2b^4}{144f}.`
\end{equation}
The PDF in this regime has the following asymptotic behavior
\begin{equation}
P(R_g^2)=\left(\frac{72f}{\pi}\right)^{1/2}\left(\frac{e\pi^2}{3}\right)^{3f/2}\frac{\alpha^{3f}}{Nb^2}e^{-\frac{\pi^2f\alpha^2}{4}}.
\end{equation}
In the opposite regime of a collapsed macromolecule, $\alpha\ll 1$, the saddle-point is $\omega_0\simeq 3i/{\alpha^2}$. Thus, one can get
\begin{equation}
W(\xi_0)\simeq \frac{9f}{4\alpha^2}+\frac{3f}{2}\ln{\frac{\alpha^2}{6}},~
W^{\prime\prime}(\xi_0)\simeq \frac{N^2b^4\alpha^6}{162f}.
\end{equation}
After the integration, we arrive at
\begin{equation}
P(R_g^2)\simeq 6^{3f/2}\left(\frac{243f}{\pi}\right)^{1/2}\frac{\alpha^{-3f}}{Nb^2}\exp\left[-\frac{9f}{4\alpha^2}\right].
\end{equation}
Collecting together the obtained results, we get
\begin{equation}
\label{PDF_2}
P(R_g^2)\simeq
\begin{cases}
\left(\frac{72f}{\pi}\right)^{1/2}\left(\frac{\pi^2e}{3}\right)^{3f/2}\frac{\alpha^{3f}}{Nb^2}\exp\left[-\frac{\pi^2f\alpha^2}{4}\right], &\alpha\gg 1\,\\
\left(\frac{243f}{\pi}\right)^{1/2}6^{3f/2}\frac{\alpha^{-3f}}{Nb^2}\exp\left[-\frac{9f}{4\alpha^2}\right], &\alpha\ll 1.
\end{cases}
\end{equation}
 
The conformational entropy in the thermodynamic limit $f\gg 1$ has the following limiting behavior
\begin{equation}
\label{entropy2}
S(R_{g}^2)/f\simeq
\begin{cases}
-\frac{\pi^2\alpha^2}{4}+3\ln{\alpha}+\frac{3}{2}\ln{\frac{\pi^2e}{3}}, &\alpha\gg 1\,\\
-\frac{9}{4\alpha^2}-3\ln{\alpha}+\frac{3}{2}\ln{6}, &\alpha\ll 1.
\end{cases}
\end{equation}

It is instructive to discuss the asymptotic behavior of the gyration radius distribution function for the Gaussian ring polymer as well. As it follows from eq. (\ref{char_2}), the characteristic function for the ring Gaussian polymer ($f=1$) is 
\begin{equation}
K(\xi)=\left(\frac{\omega}{2\sin{\frac{\omega}{2}}}\right)^3.
\end{equation}
Performing analogous calculations as below and in ref. \cite{Budkov2016}, we arrive at the following asymptotic expressions for PDF
\begin{equation}
\label{ring}
P(R_g^2)\simeq
\begin{cases}
\frac{2^{7/2}e^{1/2}3^{1/2}\pi^{11/2}}{Nb^2}\alpha^4\exp\left[-\frac{\pi^2\alpha^2}{2}\right], &\alpha\gg 1\,\\
\frac{6^5}{(2\pi)^{1/2}Nb^2}\alpha^{-9}\exp\left[-\frac{9}{2\alpha^2}\right], &\alpha\ll 1.
\end{cases}
\end{equation}

\section{Discussion of results}
\begin{figure}[h]
\centering
\includegraphics[width=0.8\linewidth]{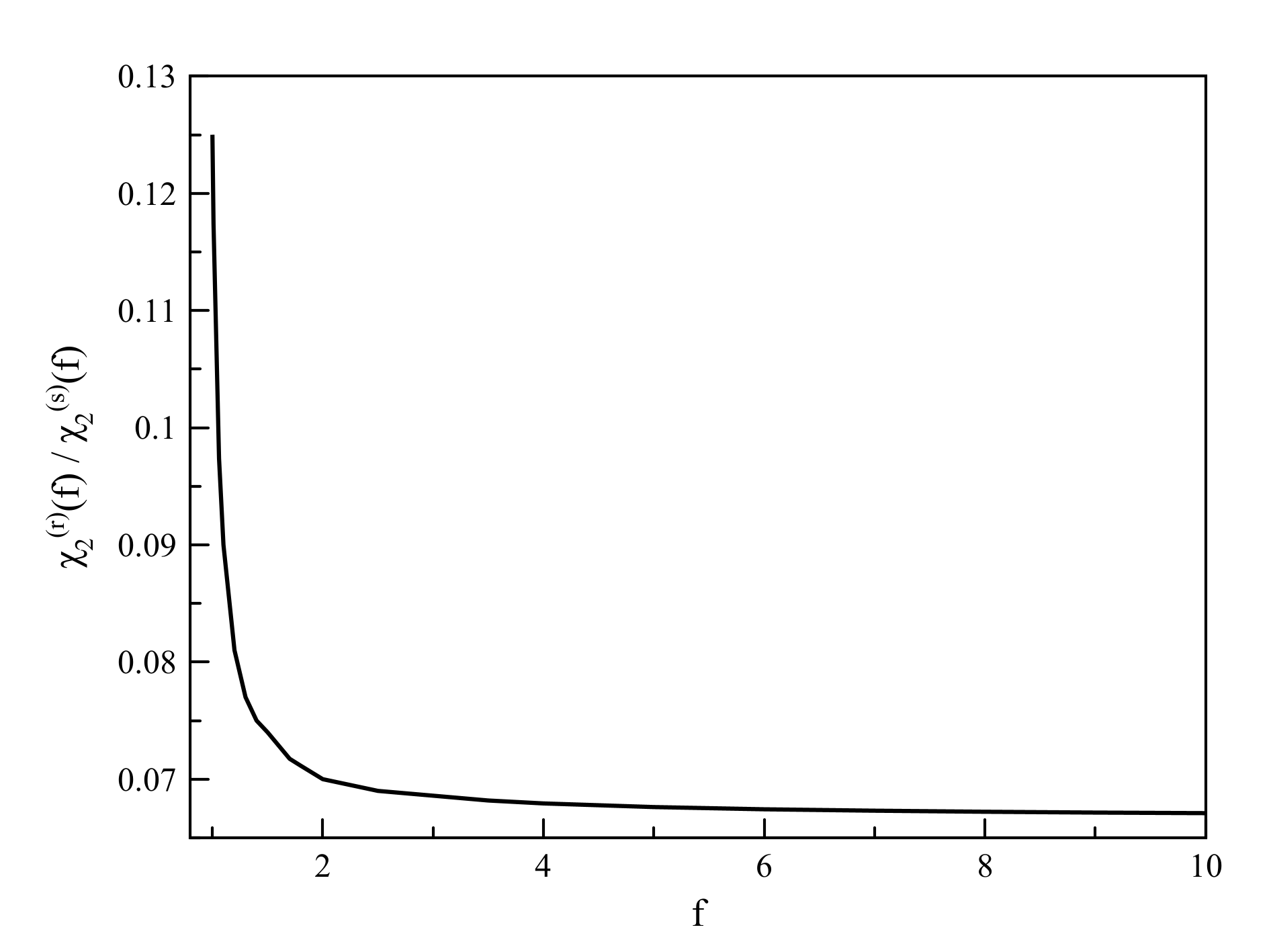}
\caption{The ratio $\chi_{2}^{(r)}(f)/\chi_{2}^{(s)}(f)$ as a function of number of arms $f$.}
\label{fig:chi} 
\end{figure}

As it follows from eqs. (\ref{PDF1}) and  (\ref{PDF_2}), the probability distribution function at $f\gg 1$ for a Gaussian rosette at $\alpha \gg 1$ behaves as $\sim \exp\left[-\pi^2f\alpha^2/4\right]$, whereas for a star polymer - as $\sim \exp\left[-3\pi^2f\alpha^2/16\right]$. Thus, the first distribution function goes to zero at $\alpha \to \infty$ much faster than the second one. At $\alpha\ll 1$, the distribution function, described by the function $\sim \exp\left[-9f/(4\alpha^2)\right]$, also tends to zero much faster than the second distribution function, which behaves as $\sim \exp\left[-3f/(4\alpha^2)\right]$. Therefore, the probability distribution function of the Gaussian rosette gyration radius is localized near its maximal value more densely than the corresponding distribution function for the Gaussian star. Thus, we can say that the Gaussian star with a large number of arms is not only larger \cite{blavatska2015conformational}, but also much softer than the Gaussian rosette. This conclusion can be made using other arguments. As is well known \cite{gnedenko2018theory}, the localization of the distribution can be quantitatively characterized by the dispersion $\chi_{2}=\mu_2-\mu_1^2$. In order to compare the localization of the gyration radius distributions in the limit $f\to \infty$, it is necessary to calculate $\lim_{f\to\infty}\chi_{2}^{(r)}(f)/\chi_{2}^{(s)}(f)$, where $\chi_{2}^{(r)}(f)$ and $\chi_{2}^{(s)}(f)$ are the dispersions of the gyration radius distributions for the rosette and the star, respectively. Using eqs. (\ref{chi}) and (\ref{chi2}) for $k=2$, we obtain 
\begin{equation}
\lim_{f\to\infty}\frac{\chi_{2}^{(r)}(f)}{\chi_{2}^{(s)}(f)}=\frac{1}{15}.  
\end{equation}
Thus, the limit of the ratio of the dispersions is much smaller than unity, which additionally confirms the above conclusion.  Fig. (\ref{fig:chi}) shows the dependence of ratio $\chi_{2}^{(r)}(f)/\chi_{2}^{(s)}(f)$ on the number $f$ of arms. As is seen, the limit $1/15$ is already achieved at $f\simeq 10$.
Note that the obtained exact values of cumulants and moments can be used as a starting point to apply the renormalization group technique to account for the volume interactions between monomer segments of nonideal star-like polymers \cite{douglas1984penetration,douglas1990characterization}.

We would also like to note that the obtained asymptotic relations for the gyration radius distributions for the Gaussian rosette (\ref{PDF1}) and single Gaussian ring polymer (\ref{ring}) have limited applicability to real macromolecules. As is well known, an important role in real ring polymers must be played by the topological constraints~\cite{khokhlov1994statistical,douglas1990characterization,koniaris1991knottedness} on possible conformations that are not taken into account in the present study. Thus, the obtained results can be used only for qualitative analysis of the conformational behavior of star-like polymers with ring constituents. We would like to note that a successful attempt to effectively take into account the topological constrains in ring polymers was recently proposed in paper \cite{polovnikov2018effective}.

\begin{figure}[h]
\centering
\includegraphics[width=0.8\linewidth]{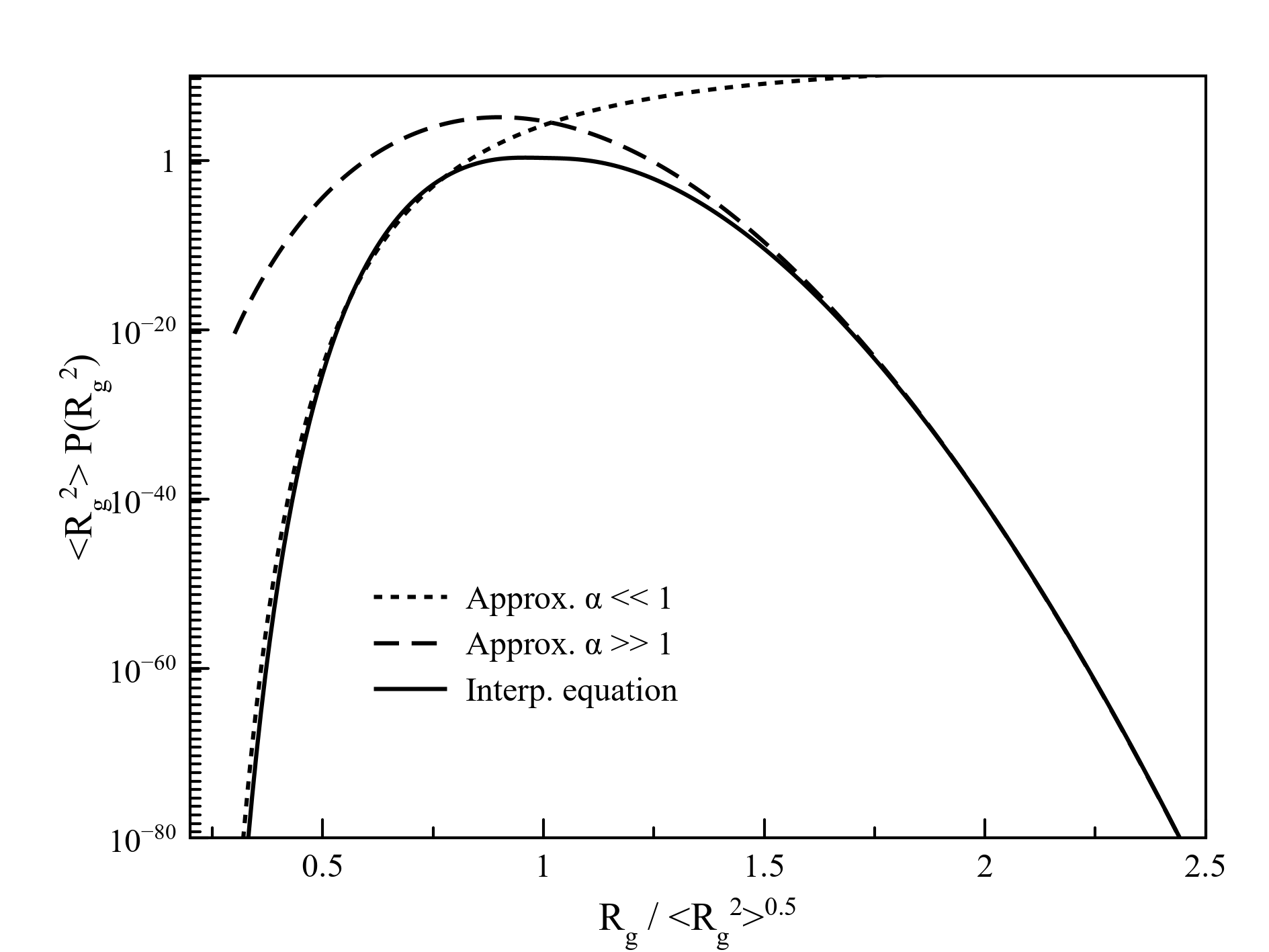}
\caption{Comparison between two limiting cases (\ref{PDF1}) of the probability distribution function for star polymer with $f = 30$ and proposed interpolation formula (\ref{eq:interpolation}).}
\label{fig:corr_pore} 
\end{figure}

Based on the two limiting expressions (\ref{PDF1}), we propose the following interpolation of the distribution function for a star polymer in the limit $f \gg 1$ and $N \gg 1$:
\begin{equation}
p(\alpha^2)=\left<R_{g}^2\right> P(R_g^2)=C_{f} \exp\left[-\frac{3f}{4\alpha^2}\right]\exp\left[-\frac{3\pi^2f\alpha^2}{16}\right]\left(\alpha^{2nf}+\alpha^{2mf}\right), \label{eq:interpolation}
\end{equation}
where $C_f$ is a normalization constant, $n$ and $m$ are the fitting parameters. Eq.(\ref{eq:interpolation}) was constructed in order to reproduce the exponential decays and the exponential function with two adjustable parameters describing the transition between them. The parameters were chosen from the least square optimization (performed via the Matlab function \textit{fminsearch}). We approximated the ratio between the first moment calculated by the interpolation formula and Eq.(\ref{eq:star_first_moment}), and the ratio between the interpolated PDF and its limiting expression at $\alpha = 4$ (the value is chosen arbitrarily and represents the "infinity" point). The resulting parameters are summarized in Table \ref{tab:I} for several different values of $f$. One can see that the value of $\bar{\mu}_2$ tends to unity as $f$ increases. The values of the adjustable parameters $n$ and $m$ seem to tend to certain saturation values as $f$ increases. Fig. \ref{fig:corr_pore} demonstrates the quality of the interpolation; one can see that the values of the interpolated PDF at $\alpha\gg 1$ coincide with the asymptotic expression more precisely than that at $\alpha\ll 1$. The discrepancy indicates the possibly erroneous assumption regarding the PDF behavior in the "middle" of the $\alpha$ range. However, it could easily turn out that the real PDF cannot be expressed through a combination of elementary functions, as it is in the case of a linear Gaussian chain \cite{yamakawa1971modern}.

\begin{table}[]
\caption{Optimized parameters of interpolation formula Eq.(\ref{eq:interpolation}) for different $f$ values; $\bar{\mu}_2=\mu_2/\left<R_{g}^2\right>^2$ is the ratio between the second moment calculated with the proposed interpolation formula and the square of (\ref{eq:star_first_moment}); $\bar{\mu}_{2}^{(theor)}$ is calculated using eqs. (\ref{eq:star_first_moment}) and (\ref{mu2}).}  
\label{tab:I}
\begin{tabular}{l|l|l|l|l|l}
\hline
$f$  & $n$      & $m$      & $C_f$              & $\bar{\mu}_2$ & $\bar{\mu}_2^{(theor)}$  \\ \hline
10 & 1.3824 & 0.5789 & $1.4707~10^{11}$ & 1.061 & 1.046  \\ \hline
15 & 1.3666 & 0.6760 & $7.724~10^{16}$  & 1.043 & 1.030 \\ \hline
20 & 1.359  & 0.7244 & $3.7884~10^{22}$ & 1.034 & 1.023 \\ \hline
25 & 1.3546 & 0.7533 & $1.7922~10^{28}$ & 1.029 & 1.018 \\ \hline
30 & 1.3510 & 0.7725 & $8.2905~10^{33}$ & 1.020  & 1.015\\ \hline
\end{tabular}
\end{table}

It should be mentioned that the interpolation formula of the probability distribution function allows us to describe the thermodynamic behavior of the star polymer. Starting from the entropy of the ideal polymeric star: 
\begin{equation}
\label{entropy}
S(R_g^2)/k_B f = -\frac{3}{4 \alpha^2} - \frac{3 \pi^2 \alpha^2}{16} + f^{-1} \ln\biggl(\alpha^{2nf}+\alpha^{2mf}\biggl) + \textit{const},
\end{equation}
where $\textit{const}$ is a constant which does not depend on $R_g$, one can calculate other thermodynamic potentials, for instance, the Helmholtz free energy \cite{rubinstein2003polymer,khokhlov1994statistical}. This fact is of interest for the developing Flory-type models \cite{budkov2017flory,budkov2018models} aiming to study conformational behavior of a polymer in a solution. We would also like to note that eq. (\ref{entropy}) can be used to describe the conformational entropy of the star-like macromolecules in the theta-solvent regime \cite{khokhlov1994statistical}.

\section{Conclusions} 
In conclusion, using the path integral approach, we have for the first time calculated the characteristic functions of the gyration radius distributions for Gaussian star and Gaussian rosette polymers. We have also been the first to obtain analytical expressions for the cumulants of both distributions. Using the steepest descent method, we have estimated the probability distribution functions of the gyration radius in the limit of a large number of arms of the star and rosette in two limiting regimes: for strongly expanded and strongly collapsed polymers. We have found that in the regime of a large gyration radius relative to its mean-square value the probability distribution functions in both cases can be described by the Gaussian functions. In the opposite regime of a strongly collapsed macromolecule both distribution functions tend to zero faster than any power of the gyration radius. Based on the asymptotic analysis, we have found that the probability distribution function for the rosette is more densely localized near its maximum than that is for the star polymer. Using the nonlinear least-square approach, we have proposed the interpolation formula for the probability distribution function of the gyration radius for the Gaussian star macromolecule.

\section*{Acknowledgments}
The research is supported by the grant from the President of the Russian
Federation (project MD-341.2021.1.3). The research was partially supported by grant 18-29-06008 from the RFBR.

\bibliography{references}
\end{document}